\documentclass[aps,pre,reprint,amsmath,amssymb]{revtex4-2}
\usepackage[pdftex]{graphicx}
\usepackage{subfigure}
\usepackage{dcolumn}
\usepackage{epstopdf}
\usepackage{color}
\usepackage{upgreek}
\usepackage[hidelinks]{hyperref}

\usepackage{bm}

\makeatletter
\def\btt#1{\texttt{\@backslashchar#1}}%
\DeclareRobustCommand\bblash{\btt{\@backslashchar}}%
\makeatother

\begin{document}

\title{Dynamics of a sheared twist bend nematic liquid crystal}

\date{\today}
\author{M. Praveen Kumar$^{1}$, Jakub Karcz$^{2}$, Przemyslaw
 Kula$^{2}$,  and Surajit Dhara$^{1}$}
\email{surajit@uohyd.ac.in} 
\affiliation{$^1$School of Physics, University of Hyderabad, Hyderabad-500046, India\\
$^{2}$Institute of Chemistry, Faculty of Advanced Technologies and Chemistry, Military University of Technology, Warsaw, Poland. }

\begin{abstract}
 We study the flow behaviour of a twist-bend nematic (N\textsubscript{TB}) liquid crystal. It shows three distinct shear stress ($\sigma$) responses in a certain range of temperatures and shear rates ($\dot{\gamma}$). In Region-I, $\sigma\sim\sqrt{\dot{\gamma}}$, in region-II, the stress shows a plateau, characterised by a power law $\sigma\sim{\dot{\gamma}}^{\alpha}$, where $\alpha\sim0.1-0.4$ and in region-III, $\sigma\sim\dot{\gamma}$. With increasing shear rate, $\sigma$ changes continuously from region-I to II, whereas it changes discontinuously with a hysteresis from region-II to III. 
 In the plateau (region-II), we observe a dynamic stress fluctuations, exhibiting regular, periodic and quasiperiodic oscillations under the application of steady shear. The observed spatiotemporal dynamics in our experiments are close to those were predicted theoretically in sheared nematogenic fluids.
  
 \end{abstract}
\preprint{HEP/123-qed}
\maketitle

\section{Introduction}

 Structure and flow behaviour of complex fluids such as colloidal suspensions, polymeric systems, surfactant gels, and liquid crystals have been a subject of increasing interests~\cite{larson,dit,iitm,fielding}. These soft materials exhibit  a strong response under moderate or weak external perturbations. Among them, liquid crystals (LCs) exhibit mesomorphic properties, i.e., variety of thermodynamically stable phases. Some commonly occurring phases in thermotropic LCs are nematic, smectic and cholesteric \cite{pg,ch}. The nematic phase has only orientational order,  smectic has additionally a lamellar structure and in cholesteric the director (average orientation direction of the molecules) forms a helical structure. Because of the weak intermolecular interactions, the structure of LCs can be easily deformed by external forces such as electric, magnetic \cite{blinov} and stress field \cite{safina,cates,sd,nagita,mr,rm,ad}. In fact the optical response of nematic LCs due to the electric field has been exploited in the liquid crystal display (LCDs) applications. The applied stress field not only changes the orientation and equilibrium structure but also changes the flow behaviour of the LCs which are mostly unknown for many new LCs.
 

\begin{figure}[!ht]
\center\includegraphics[scale=0.6]{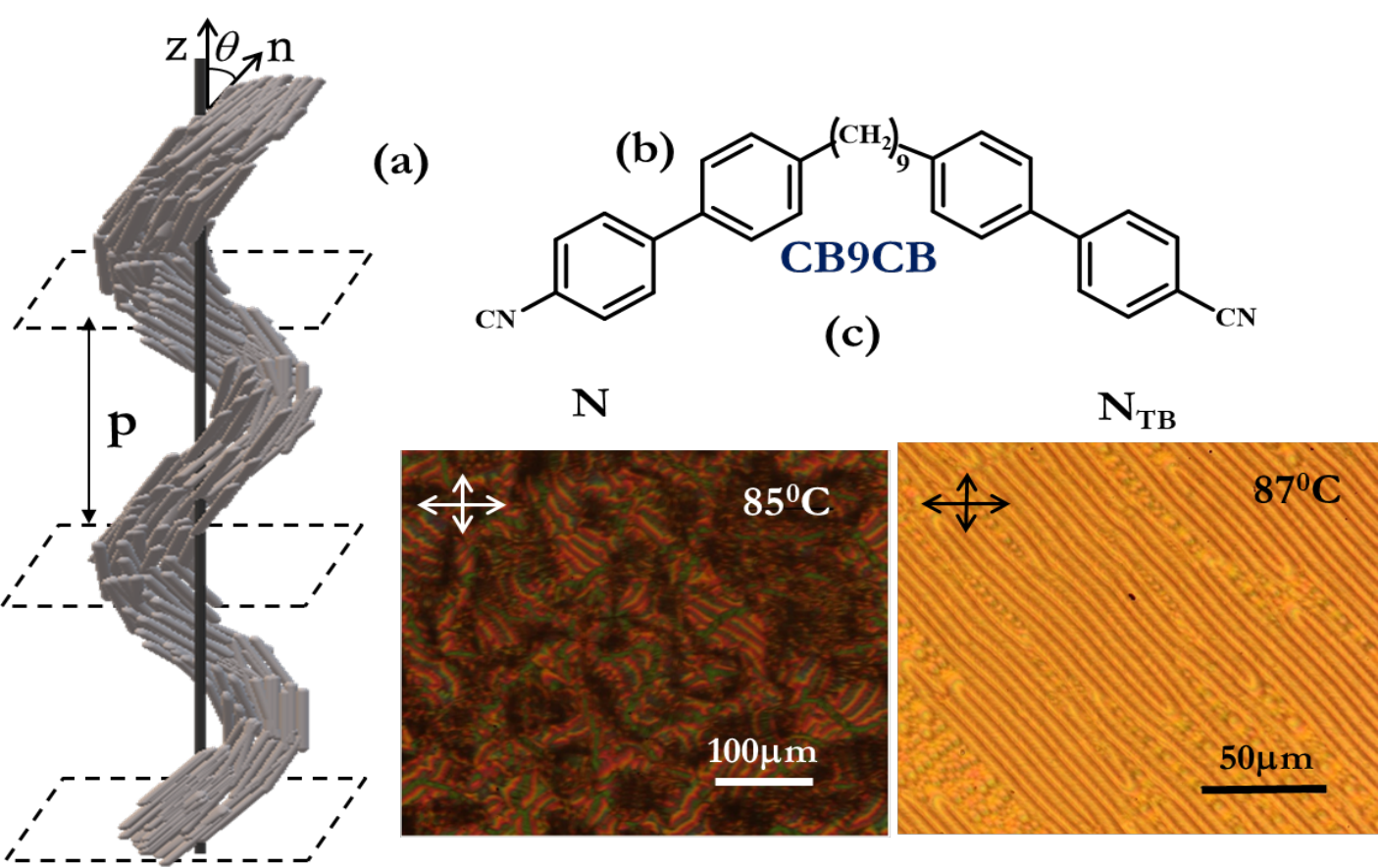}
\caption{(a) Schematic view of heliconical  orientation of the molecules in the N\textsubscript{TB} phase. Separation between the two dotted planes represents pitch $p$, equivalent to the pseudolayer thickness. (b) Chemical structure of the compound CB9CB. (c) Polarizing optical microscope textures  in the N\textsubscript{TB} phase in a homeotropic (left) and in a homogeneous (right) cells \label{fig:figure1}.}
\end{figure}

In the recent past, a new nematic phase, called the nematic twist bend (N\textsubscript{TB}), made of achiral molecules has been discovered, in which the director forms an oblique helicoid ($0<\theta<\pi/2$) with a nanoscale periodicity (Fig.\ref{fig:figure1}(a))~\cite{id,vp,vb,mco,chd,mcl,prx,oleg1,noel}. This is very different than the structure of cholesteric LCs made of chiral molecules, where $\theta=\pi/2$ and the pitch is much larger  (a few hundred nm) \cite{ch}.  The heliconical pitch of N\textsubscript{TB} is of the order of a few molecular length and can be regarded as ``pseudo-layers" without a true mass density wave unlike smectics \cite{cm}.  Interestingly, the pseudo-layers of the N\textsubscript{TB} LCs offer compression elastic modulus, similar to the conventional smectics but with a lower magnitude~\cite{prm,sms}. It is important to ask how do this novel nematic phase respond to the applied stress field. Is the flow behaviour of N\textsubscript{TB} LCs, a nematic-like or a smectic-like ? To address these questions, we systematically studied the flow behaviour of a N\textsubscript{TB} LC. We show that the studied  N\textsubscript{TB} LC exhibits a shear induced dynamic first order transition in a certain temperature and shear-rate range. In the intermediate shear-rate range it shows a stress oscillations which proceeds from periodic to aperiodic oscillations with increasing temperature.

\section{Experiment}
We have synthesized a dimeric liquid crystal, $\alpha$, $\omega$-bis(4,4-$^{'}$cyanobiphenyl nonane), which is known as CB9CB in short. A summary of the synthesis scheme is presented in Appendix-B. Two cyanobiphenyl units are connected through nine ($n=9$) flexible methylene units (Fig.\ref{fig:figure1}(b)). The phase transitions and textures were observed using polarizing optical microscope (Olympus BX51) and a temperature controller (Mettler FP 90). In cooling, it exhibits the following phase transitions: I $124^\circ$C N $108^\circ$C N\textsubscript{TB} $84^\circ$C Cr. These transition temperatures are similar to those reported by Paterson \textit{et al.}\cite{da}. Polarising optical microscope textures in a planar and in a homeotropic cells are shown in Fig.\ref{fig:figure1}(c). The textures of N\textsubscript{TB} phase are somewhat similar to those of the focal conic textures of usual SmA LCs. We have used a strain controlled Rheometer (MCR 501, Anton Paar) with a cone-plate measuring system having plate diameter of 25 mm and the cone angle of $1^\circ$. A Peltier temperature controller was attached with the bottom plate for controlling the temperature with an accuracy of $0.1^\circ$C. A hood was used to cover the measuring system for maintaining uniformity of the sample temperature. Fresh sample was mounted for each measurement and it was heated above the nematic-isotropic transition temperature and then cooled to the N\textsubscript{TB} phase and presheared for 5 minutes at a low shear rate (5 s\textsuperscript{-1}) before starting the measurements. The phase transition temperatures of the mounted samples in the Rheometer were ascertained from the temperature-dependent viscosity that shows characteristic changes at the phase transitions. The stress relaxation data is collected at each second for the duration of 60 minutes.

\section{Results and discussion}

\begin{figure}[!ht]
\center\includegraphics[scale=0.6]{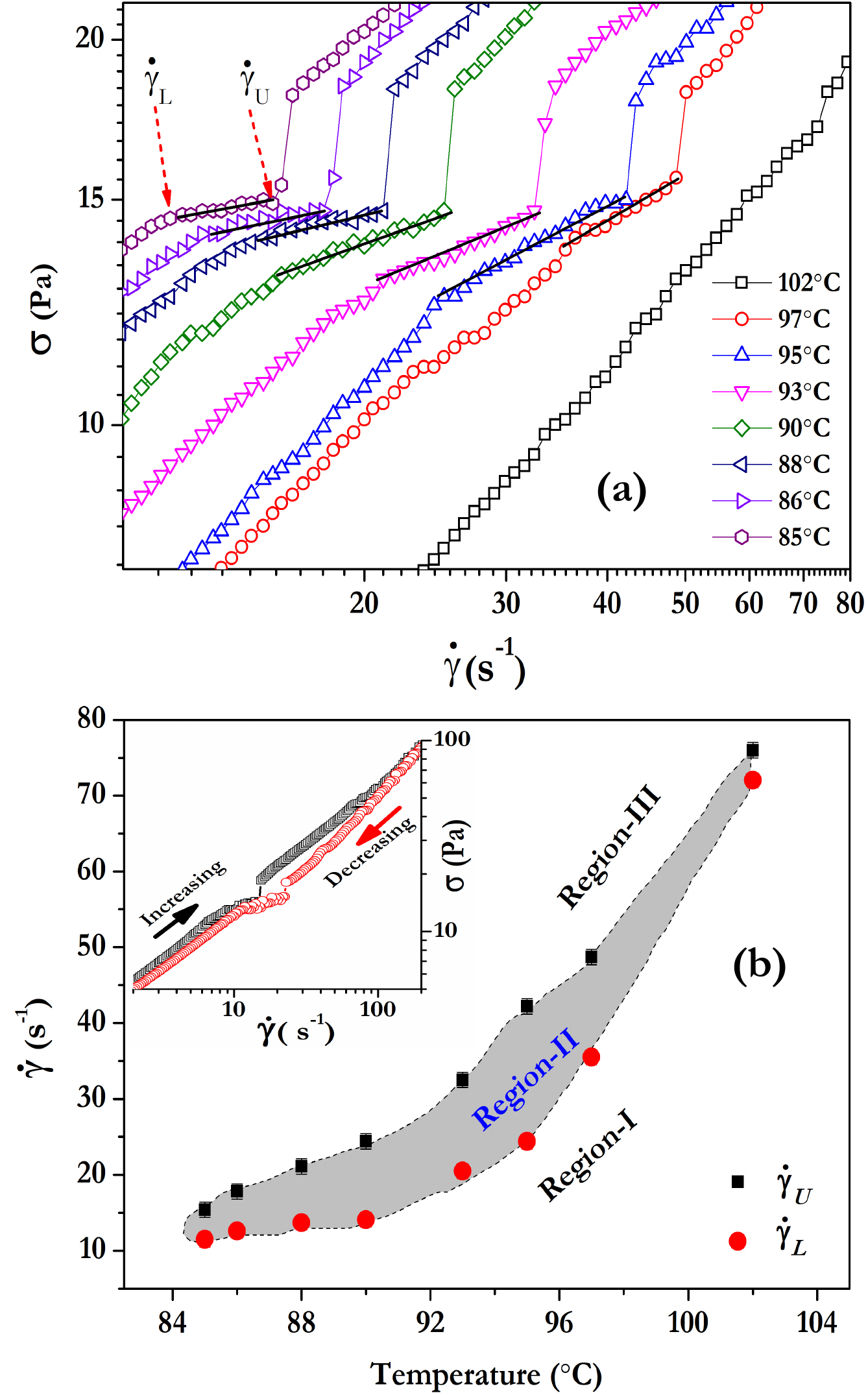}
\caption{(a) Flow curves at different temperatures in the N\textsubscript{TB} phase of CB9CB LC. Solid lines show best fits to the scaling relation $\sigma\sim{\dot{\gamma}}^{\alpha}$. (b) Temperature variation of the onset shear rates ${\dot{\gamma}}_{L}$ (red circles) and the terminal shear rates ${\dot{\gamma}}_{U}$ (black squares) of the stress plateaus. Dotted curve is a guide to the eye. Inset shows the thermal hysteresis of the transition from region-II to the region-III at temperature T=85$^\circ$C. The difference between the two shear rates in heating and cooling at the transition is $\Delta\dot{\gamma}=8$ s\textsuperscript{-1}.
\label{fig:figure2} }
\end{figure}

Figure \ref{fig:figure2}(a) shows the flow curves at different temperatures in the N\textsubscript{TB} phase in log-log scale. The data in the linear scale from the starting shear rate is presented in Appendix-A (Fig.\ref{fig:figure8}). At a fixed temperature, initially the stress ($\sigma$) increases with shear rate ($\dot{\gamma}$) and at a particular shear rate ($\dot{\gamma_L}$), the stress changes slope and tends to saturate, showing a stress-plateau. At a certain higher shear rate ($\dot{\gamma_U}$), the stress jumps discontinuously to a larger value from the plateau and continue to increase with increasing shear rate. Similar discontinuous transition is also observed with decreasing shear rate. But the transition takes place at a slightly higher shear rates, showing a clear thermal hysteresis [inset to Fig.\ref{fig:figure2}(b)], a typical characteristic of a discontinuous transition.   
 With increasing temperature the discontinuous transition takes place at a higher shear rates. The variations of the shear rates $\dot{\gamma_L}$ and $\dot{\gamma_U}$ with temperature are shown in Fig.\ref{fig:figure2}(b). We observe three distinct regions in the $\dot{\gamma}$-T plane. The transition from the region-I to the region-II takes place continuously whereas from the region-II to the region-III, the transition takes place discontinuously. Beyond a certain temperature both the transitions are not observed ($\simeq 103$ $^\circ$C). 

\begin{figure}[htbp]
\includegraphics[scale=0.6]{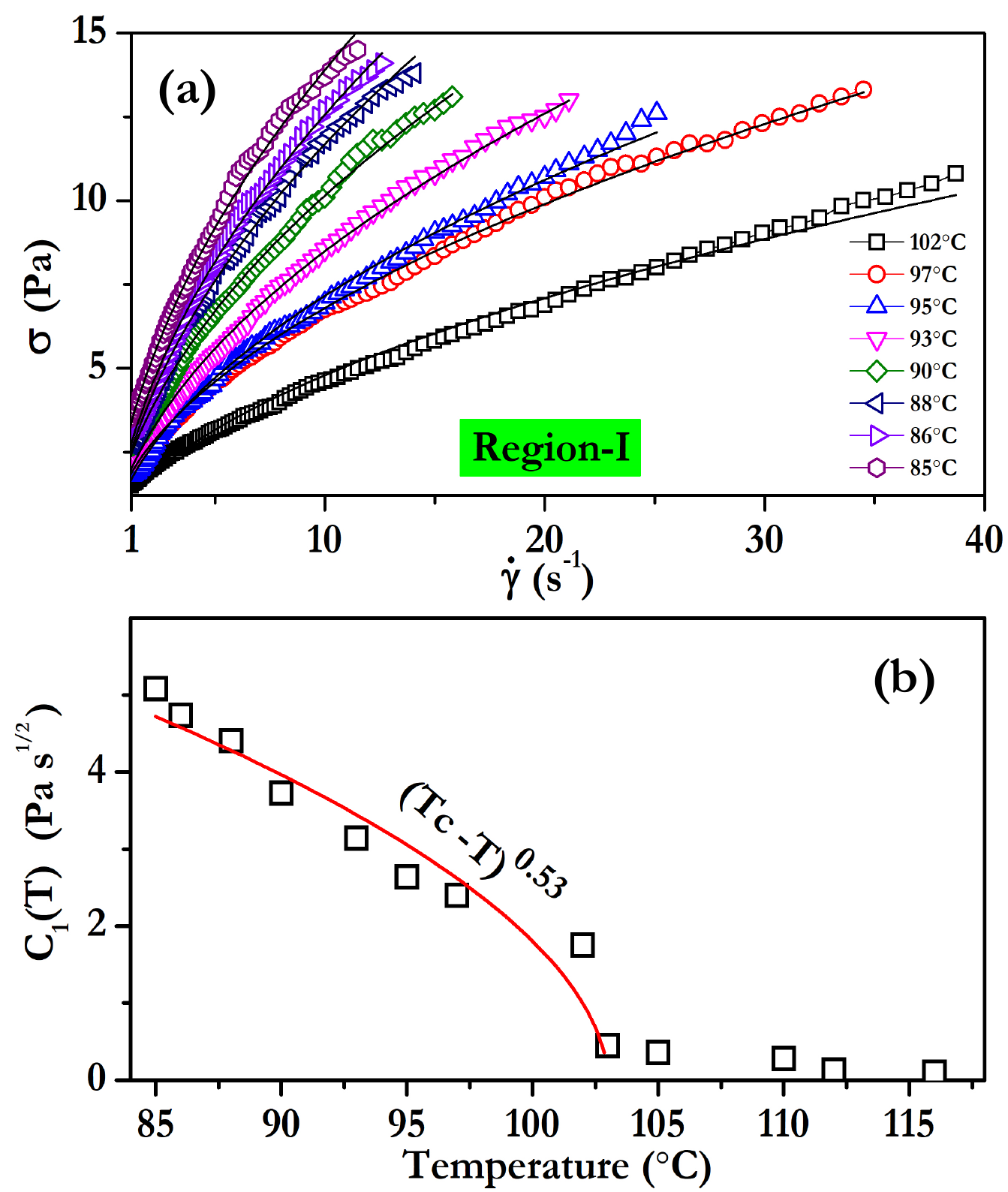}
\center\caption{ (a) Variation of stress $\sigma$ with $\sqrt{\dot{\gamma}}$ in region 1 at different temperatures in the N\textsubscript{TB} phase. Solid lines are best fits to the scaling relation $\sigma \sim C_1(T)\sqrt{\dot{\gamma}}$. (b) Temperature variation of slope $C_1(T)$. Solid red curve is a least square fit to $C_1(T)\sim(T_c-T)^{0.53\pm 0.01}$ with $T_c=103^\circ$C.
\label{fig:figure3}}
\end{figure}

Figure \ref{fig:figure3}(a) shows the stress versus the shear rate at various temperatures in region-I. It shows a Non-Newtonian flow behaviour and the data can be fitted to the scaling relation $\sigma\sim C_{1}(T){\dot{\gamma}^{1/2}}$, where $C_1(T)$ is a proportionality constant that varies with temperature. This also means that the shear viscosity decreases as $\eta\sim{\dot{\gamma}^{-1/2}}$, exhibiting a shear thinning behaviour. The constant $C_{1}(T)$ decreases with temperature and it can be fitted to $C_{1}(T)\sim(T_c-T)^{0.53\pm 0.01}$, where $T_c$ (=103 $^{\circ}$C) is the temperature above which the discontinuous transition is not detectable. It is noticed that $T_c$ is about 5$^{\circ}$C less than the nematic to N\textsubscript{TB} transition temperature (108 $^{\circ}$C).

\begin{figure}[!ht]
\center\includegraphics[scale=0.6]{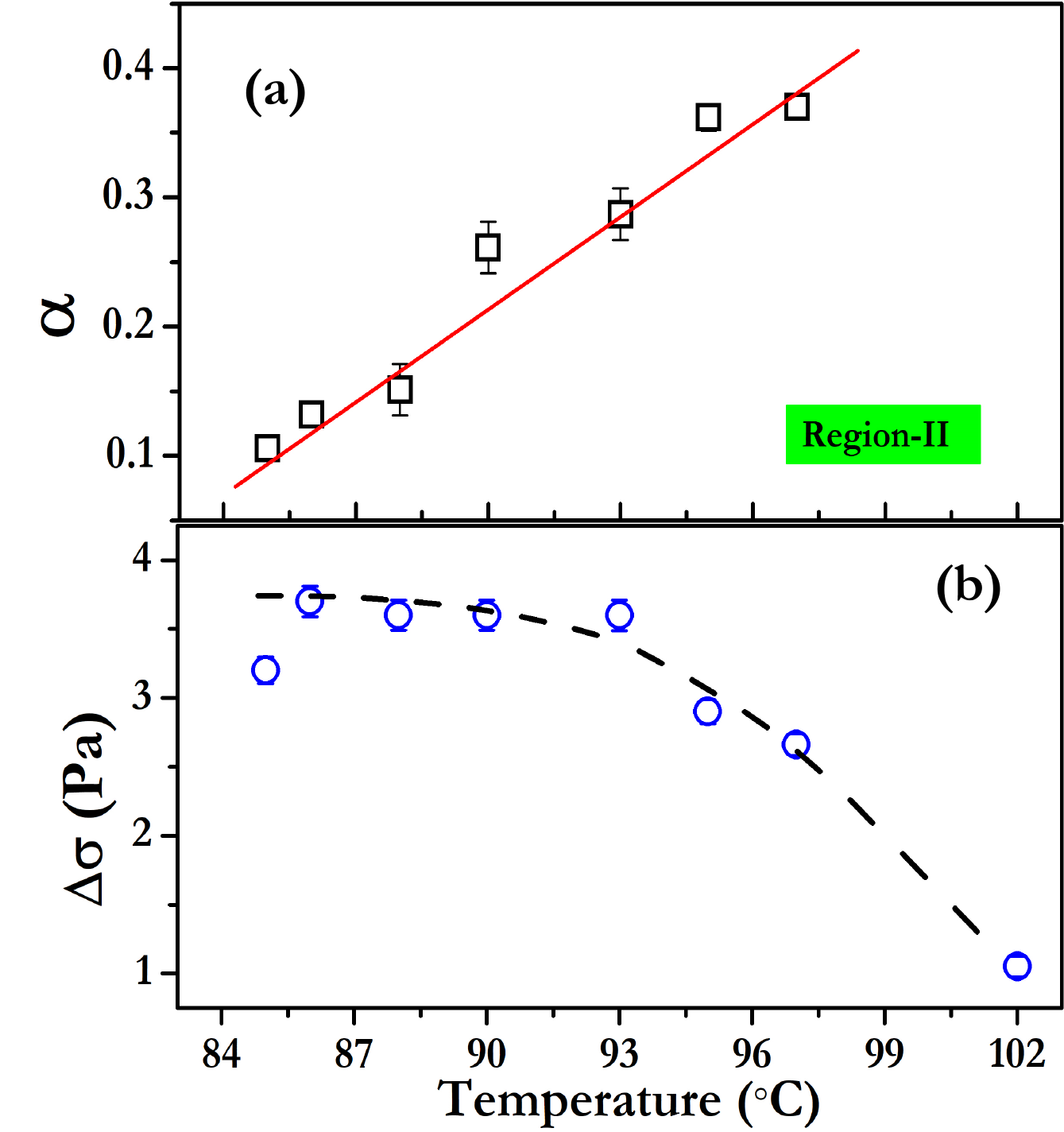}
\caption{  Temperature variation of (a) the exponent $\alpha$ of the stress plateaus in region-II. Solid red line is a best fit to the data. (b) Temperature variation of the discontinuous stress jump $\Delta\sigma$ in the N\textsubscript{TB} phase. Dashed curve is drawn as a guide to the eye.
\label{fig:figure4}}
\end{figure}

In region-II, the stress plateaus have a finite slope and they can be fitted to a power law: $\sigma\sim\dot{\gamma}^{\alpha}$ where $\alpha$ is an exponent that increases linearly with the temperature from 0.1 to 0.4 as shown in Fig.\ref{fig:figure4}(a). Similar stress plateaus have been observed in shear-banding micellar aqueous solutions of surfactant (cetyl trimethylammonium tosulate (CTAT))as a function of salt concentration\cite{rg}. It was explained, taking into account the effect of coupling of flow-concentration \cite{OD,dr,pza,rb,jb,rg,jfb,pramana,smf,gp,rbprl} and flow-microstructures\cite{smfprl}. Ours is a single component thermotropic LC hence the possibility of flow-concentration coupling is ruled out. The stress plateau in our system could be due to a shear-induced state of coexistence of N and N\textsubscript{TB} phases, an effect analogous to the shear-induced nematic-isotropic transition in wormlike micelles, wherein the low and high viscous regions are separated, forming shear-bands\cite{jb}. In this scenario the low viscous usual nematic domains are expected to appear near the top plate and the high viscous N\textsubscript{TB} domains near the bottom plate. 
At a particular shear rate ${\dot{\gamma}}_{U}$, the stress jumps discontinuously to a higher value from region-II to region-III. Figure \ref{fig:figure4}(b) shows the variation of the stress jump $\Delta\sigma$ with temperature. $\Delta\sigma$ decreases approximately from 3.5 to 1 near the critical temperature $T_c=103$ $^\circ$C. 

\begin{figure}[htbp]
\center\includegraphics[scale=0.55]{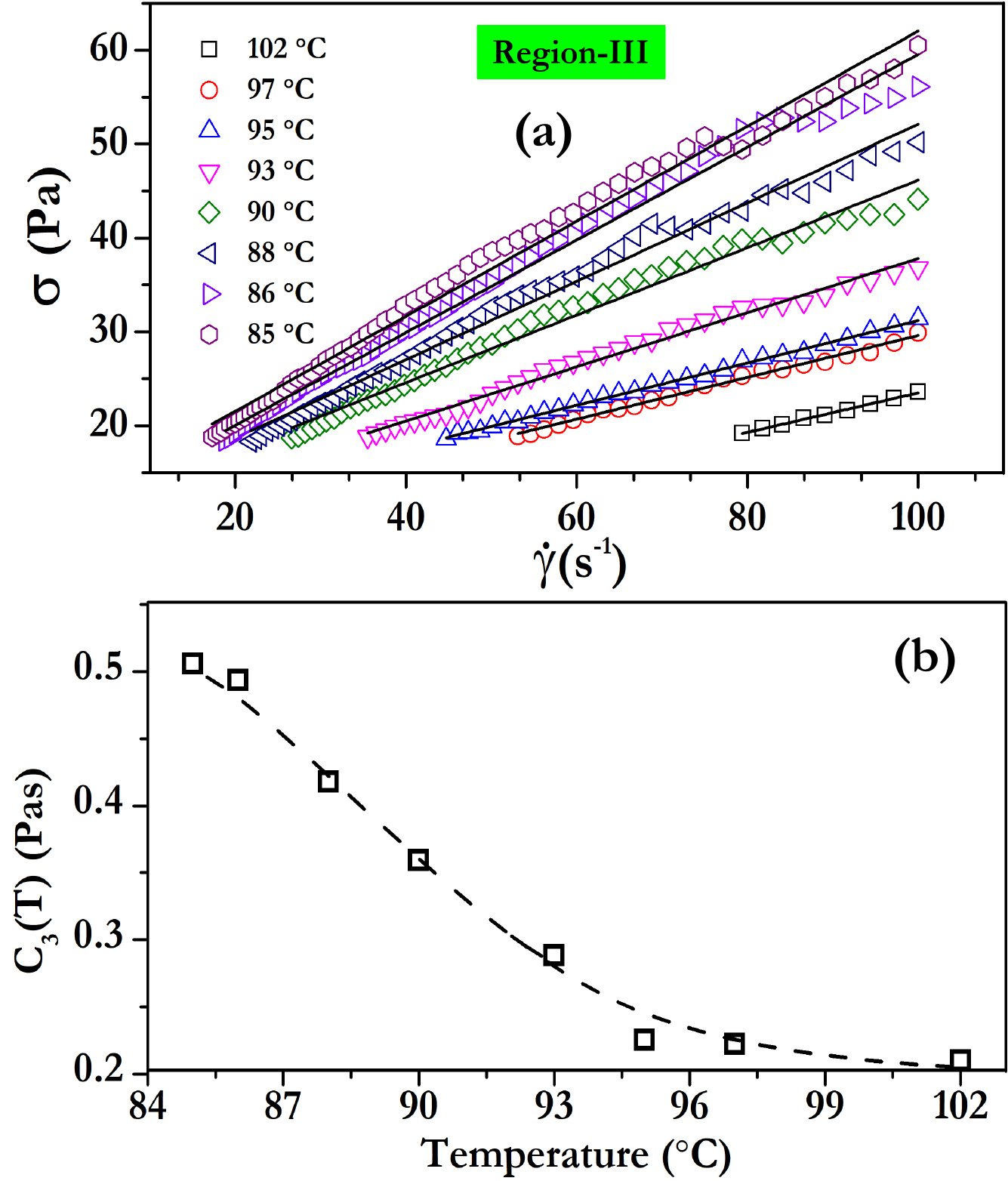}
\caption{(a) Variation of stress $\sigma$ with $\dot{\gamma}$ in region-III at different temperatures in the N\textsubscript{TB} phase. Solid lines are  best fits to the scaling relation $\sigma \sim C_3(T)\dot{\gamma}$. (b) Temperature variation of $C_3(T)$. Dashed curve is drawn as a guide to the eye.
\label{fig:figure5}}
\end{figure}

In region-III, the N\textsubscript{TB} phase shows a Newtonian flow behaviour (Fig.\ref{fig:figure5}(a)) and the stress can be fitted to the scaling relation $\sigma\sim C_3(T){\dot{\gamma}}$, where $C_3(T)$ is equivalent to the effective viscosity $\eta_{eff}$. Figure \ref{fig:figure5}(b) shows that $C_3(T)$ decreases with increasing temperature as expected. 

The effect of shear in the SmA phase of 4-cyano-4\textsuperscript{'}-octylbiphenyl (8CB) liquid crystal has been studied using a stress controlled rheometer, in which the stress was applied and the strain was measured\cite{pza}. Interestingly, the shear responses of the N\textsubscript{TB} phase of CB9CB are somewhat similar to those of the SmA phase of 8CB LC, except the effects here are seen at much higher shear rates. From the simultaneous x-ray scattering investigations of the  8CB LC, two steady state orientations of lamellae were identified in a certain temperature range of the SmA phase. In the low shear rate regime (equivalent to region-I), multilamellar cylinders are formed which are oriented along the flow direction and the system shows a non-Newtonian flow behaviour namely, $\sigma\sim\sqrt{\dot{\gamma}}$, similar to that is observed in the present sample. At high shear rate regime (equivalent to region-III), SmA layers are oriented perpendicular to the shear plane and $\sigma\sim\dot{\gamma}$. In the intermediate shear rate range, a diphasic region was observed where these two states coexist.  
Based on the overall similarities on the shear responses of the SmA and N\textsubscript{TB} phases, two possible scenarios of the N\textsubscript{TB} phase could be considered. The  stress plateau of the N\textsubscript{TB} phase could be due to the shear banding or it could be due to the coexistence of multilamellar cylinders and orientated SmA planes forming a diphasic region. The SmA phase of 8CB LC has a layer thickness $\sim 2$ nm with a true mass density wave, whereas the N\textsubscript{TB} phase of CB9CB LC has a pseudolayer structure (pitch $\sim 9$ nm) without a true mass density wave \cite{zhu}. Moreover, the pseudolayers of N\textsubscript{TB} phase offers compression elastic modulus whose magnitude is one or two orders of magnitude lower than that of the SmA phase \cite{prm}. Considering the structural similarities between the SmA and N\textsubscript{TB} phases, it is more likely that in region-I of the N\textsubscript{TB} phase, the pseudolayers form multilamellar cylinders which are oriented along the flow direction and in region-III the pseudolayers are oriented perpendicular to the shearing plates. Region-II is diphasic region wherein the nematic and  N\textsubscript{TB} coexists. However, confirmation of such dynamic features require further rheo-microscopic or rheo x-ray investigations.

\begin{figure}[htbp]
\center\includegraphics[scale=0.65]{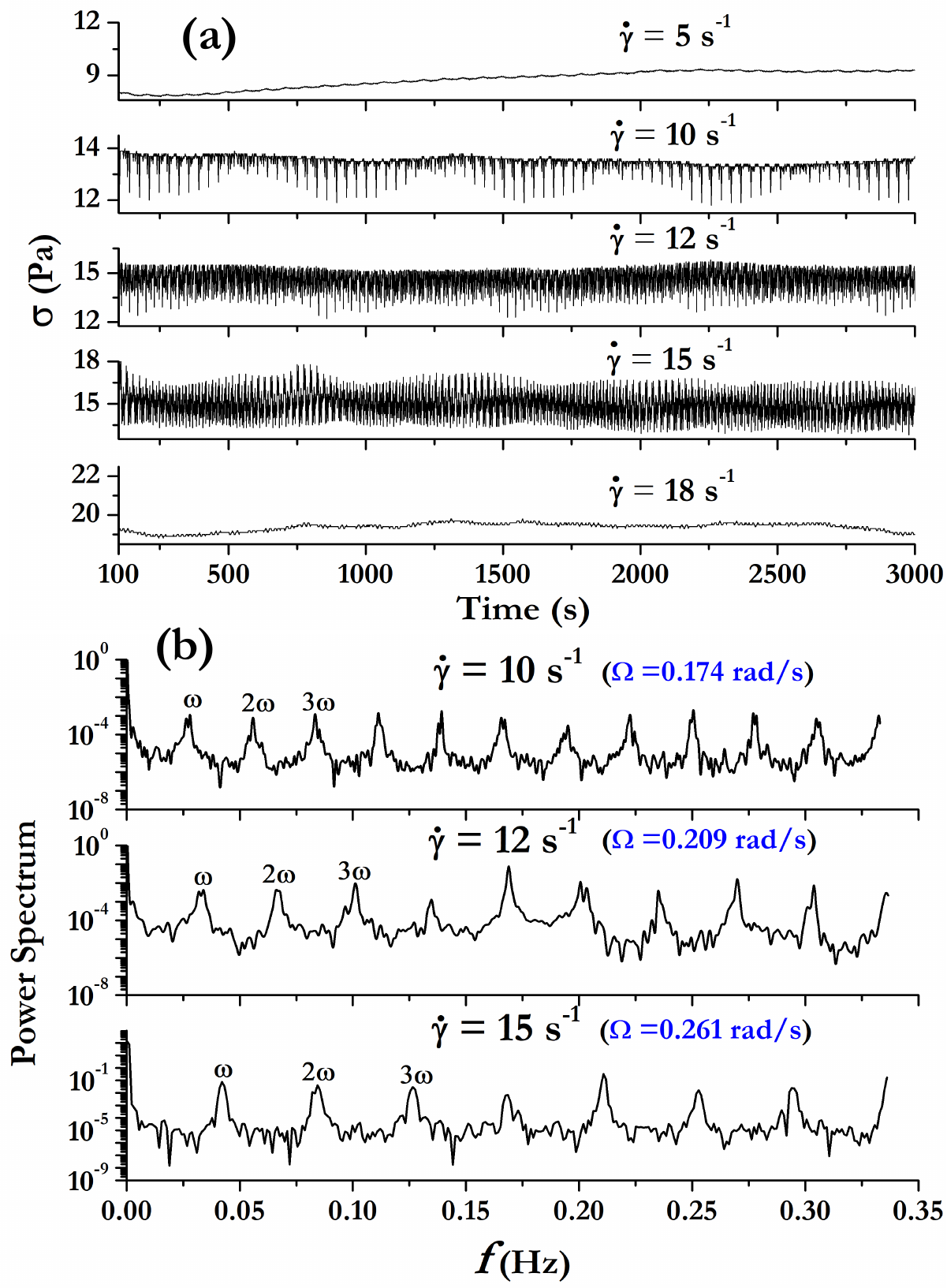}
\caption{ (a) The stress oscillation time series at a fixed temperature $85^\circ$C and different shear rates. Shear rates $\dot{\gamma}=10,12,15$ s\textsuperscript{-1} lies within the stress plateau. $\dot{\gamma}=5$ and 18 s\textsuperscript{-1} lies just below and above the stress plateau, respectively (see Fig. \ref{fig:figure2}(a)). (b) Corresponding Fourier power spectrum. A few harmonics of the fundamental frequency $\omega$ ($=2\pi f$) are labelled. Shearing frequencies calculated using $\Omega=\dot{\gamma} \tan(\alpha)$ are also shows within brackets next to the shear rates.
\label{fig:figure6}}
\end{figure}


The existence of stress plateau is often a signature of shear-banded inhomogeneous stationary flows resulting mechanical instability~\cite{gp}. The time-dependent stress in this region usually shows dynamic relaxation~\cite{rbprl,rgprl}.
In order to study the stress dynamics in the plateau region we have measured the time dependence of stress under a steady shear at a few temperatures. Figure \ref{fig:figure6}(a) shows the time-dependent stress profiles at some selected shear rates, namely $\dot{\gamma}=5,10,12,15,18$ s\textsuperscript{-1} and at a fixed temperature (T= 85$^\circ$C) in the N\textsubscript{TB} phase. Among these shear rates, 5 s\textsuperscript{-1} lies just below and shear rate 18 s\textsuperscript{-1} lies just above the plateau region. The stress values ($\sigma$) for these two shear rates (out side the plateau region) is almost constant. When the shear rate is increased to 10 s\textsuperscript{-1}, $\sigma$ shows oscillation which looks periodic and the patterns change at higher shear rates  ( see $\dot{\gamma}=12$ and 15 s\textsuperscript{-1}). The amplitude of the oscillation is much larger than the limiting stress value of the plate-cone geometry (24 mPa.s\textsuperscript{-1}), therefore, the stress signal is due to the dynamic change in the sample.
The Fourier power spectrums of the signals for the shear rates, $\dot{\gamma}=10,12$ and 15 s\textsuperscript{-1} are shown in Fig. \ref{fig:figure6}(b). The stress oscillations at 10 s\textsuperscript{-1} shows a fundamental mode ($\omega$) and its higher harmonics. The frequency $\omega$ of the fundamental mode is shifted to higher values with increasing shear rate. For example, the frequencies ($\omega=2\pi f$) for the fundamental modes for shear rates, $\dot{\gamma}=10,12$ and 15 s\textsuperscript{-1} are 0.170 rad/s, 0.205 rad/s and 0.265 rad/s, respectively. 
The driving angular frequency ($\Omega$) of the plate-cone system for a given shear rate $\dot{\gamma}$, is given by $\Omega=\dot{\gamma}\tan(\alpha)$, where $\alpha$ (=1$^{\circ}$) is the cone angle. The calculated driving frequencies $\Omega$ for the shear rates 10, 12 and 15 s\textsuperscript{-1} are 0.174, 0.209 and 0.261 rad/s, respectively which are very close to the respective fundamental frequencies i.e., $\Omega\simeq\omega$ (Fig.\ref{fig:figure6}(b)). This suggests that the stress oscillations in the plateau region at this temperature is driven by the shear force. 
\begin{figure}[htbp]
\center\includegraphics[scale=0.65]{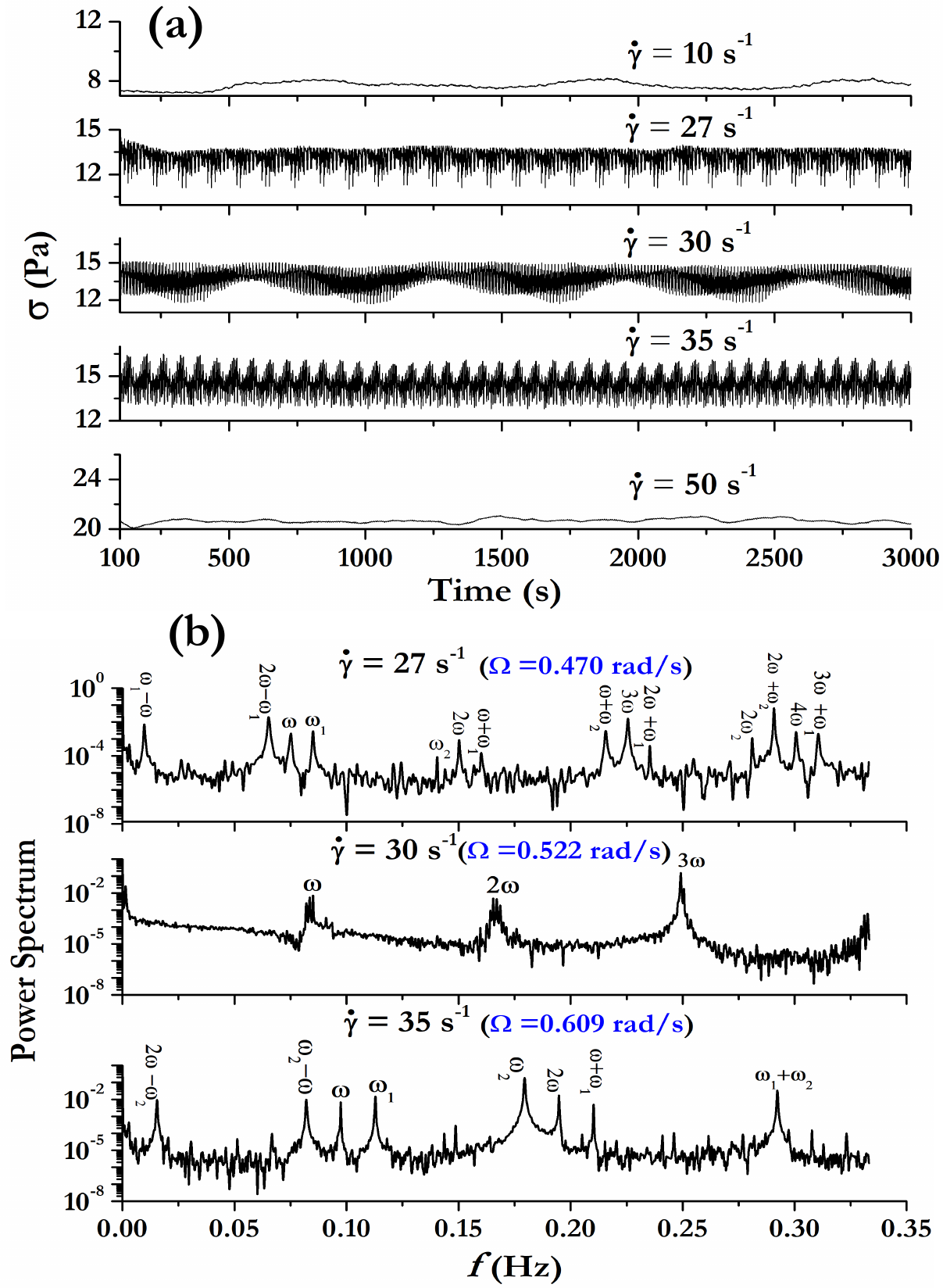}
\caption{(a) The stress oscillation time series at a fixed temperature $95^\circ$C and different shear rates. $\dot{\gamma}=27, 30, 35$ s\textsuperscript{-1} lies within the stress plateau. $\dot{\gamma}=10$ and 50 \textsuperscript{s} lies below and above the stress plateau, respectively  (see Fig. \ref{fig:figure2}(a)). (b) Corresponding Fourier power spectrum. A few harmonics of the fundamental frequency $\omega$ ($=2\pi f$) are labelled. Shearing frequencies calculated using $\Omega=\dot{\gamma} \tan(\alpha)$ are shown within brackets next to the shear rates.
\label{fig:figure7}}
\end{figure}

Figure \ref{fig:figure7} shows time-dependent stress profile and the corresponding power spectra at a higher temperature (T=95$^\circ$C). Shear rates 10 and 50 s\textsuperscript{-1} lie below and above the plateau-range and the corresponding stress values are almost constant (Fig. \ref{fig:figure7}(a)). Shear rates 27, 30 and 35 s\textsuperscript{-1} are well within the range of stress plateau. The stress signal corresponding to $\dot{\gamma}=27$ s\textsuperscript{-1} looks periodic with two modes but the power spectrum shows that there is a fundamental frequency $\omega$, which is identified as equal to the shearing frequency $\Omega$ and also its higher harmonics e.g., $2\omega$, $3\omega$ and $4\omega$. Apart from these, there are two more frequencies $\omega_1$, $\omega_2$ and their higher harmonics, and there are also several linear combinations of frequencies such as $\omega+\omega_1$, $\omega+\omega_2$, $\omega_1+2\omega$, $2\omega+\omega_2$, $3\omega+\omega_1$, $2\omega-\omega_1$ and $\omega_1-\omega$ etc. These additional features are the hallmark of quasiperiodic signal. The stress oscillation  corresponding to the shear rate 30 s\textsuperscript{-1} apparently looks periodic (Fig. \ref{fig:figure7}(a)). However, the power spectrum shows that the fundamental as well as the higher harmonics signals split within a narrow range (Fig. \ref{fig:figure7}(b)). The stress oscillation at shear rate 35 s\textsuperscript{-1} and the corresponding power spectra shows that the oscillation is quasiperiodic and apart from the three primary frequencies ($\omega$, $\omega_1$, $\omega_2$), there are linear combination of frequencies such as  $\omega+\omega_1$, $\omega_1+\omega_2$, $\omega_2-2\omega_1$ and $\omega-\omega_2$. We also performed experiments at a few higher temperatures and the overall the stress response was quasiperiodic. 

Theoretically dynamics of shared nematogenic fluids considering the coupling of the order parameter to flow have been studied based on the relaxation equation for the alignment tensor~\cite{srprl,cd1,mdas,dc1}. In certain parameter space, where flow alignment does not occur it was shown that the dynamic responses are due to the spatiotemporal fluctuations of the stress~\cite{srprl}. Also a variety of symmetry breaking transient states with out-of-plane director configuration showing complicated periodic or regular chaotic states have been predicted~\cite{rgm}. Recently nematic hydrodynamic incorporating spatial inhomogeneity under controlled shear rate and shear stress has been studied numerically\cite{cd1}. For a certain range of tumbling parameter the model predicts, three distinct states or phases, namely periodic, spatiotemporally chaotic and aligned. 
 In the N\textsubscript{TB} phase, we observe some of the features of the stress dynamics predicted in the above references. Therefore, the set of parameters used in the theory that predicts spatiotemporal dynamics in a nematic LC may correspond to our experiments.  The similarity of our experimental results with those of the theory is exciting and it encourages for further experimental work using rheo-optical and small-angle neutron scattering techniques. 


\section{Conclusion} 
In conclusion we have studied the effect of steady shear on the dynamics of a twist bend nematic liquid crystal. In the $\dot{\gamma}$-T diagram the N\textsubscript{TB} shows a continuous transition from region-I to region-II and a discontinuous transition from region-II to region-III with finite stress jump. With increasing temperature, the stress jump decreases and the transition is not detectable beyond a critical temperature, which is much below the N\textsubscript{TB} to nematic transition. The flow behaviour in region-I and region-II is nonlinear with characteristic exponents. However, in region-III the flow behaviour is linear. The overall flow behaviour in region-I and region-III are similar to that of the SmA phase of 8CB LCs. Thus, under shear, the pseudolayers of N\textsubscript{TB} responds like a usual SmA layers with a true mass density wave. The stress is plateaued in region-II, similar to that is usually observed in shear banding fluids. In the plateau region we observed both regular periodic and quasi-periodic stress oscillations which are somewhat similar to those were reported in wormlike micellar systems. It appears that with increasing temperature our system moves from regular periodic to quasi-periodic state.  Our experiments show some striking similarity with the dynamic stress response of nematogenic fluids predicted recently. 
Our study shows that the flow dynamics of the nematic twist bend liquid crystals are promising for new effects and it requires further experimental and theoretical studies.

\section{Acknowledgments}
\textbf{Acknowledgments}: SD acknowledges financial support from SERB (Ref. No:CRG/2019/000425). MPK acknowledges UGC-CSIR for fellowship. We thank Dr. Kabir Ramola from TCIS-TIFR for useful discussion.

\subsection*{\bf{APPENDIX: A}}

\begin{figure}[htbp]
\center\includegraphics[scale=0.47]{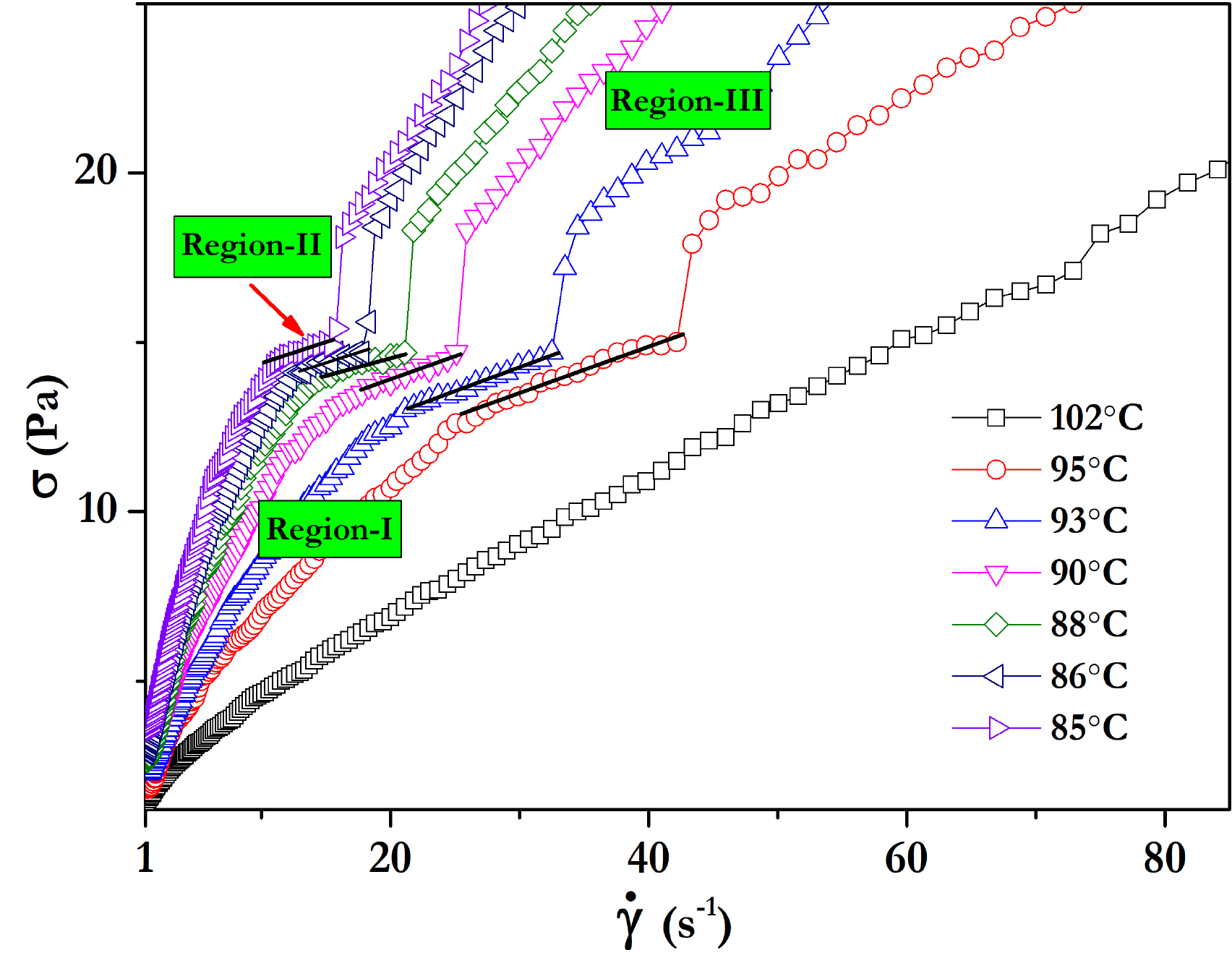}
\caption{ Shear rate dependent shear stress in the linear scale from $\dot{\gamma}=1 s^{-1}$ to 85 $s^{-1}$ at different temperatures in the N\textsubscript{TB} phase.
\label{fig:figure8}}
\end{figure}

Figure \ref{fig:figure8} shows the variation of stress $\sigma$ with shear rate $\dot{\gamma}$ in the linear scale. Different regions described in the main text are marked.  

\subsection*{\bf{APPENDIX: B}}

\begin{figure}[htbp]
\center\includegraphics[scale=0.5]{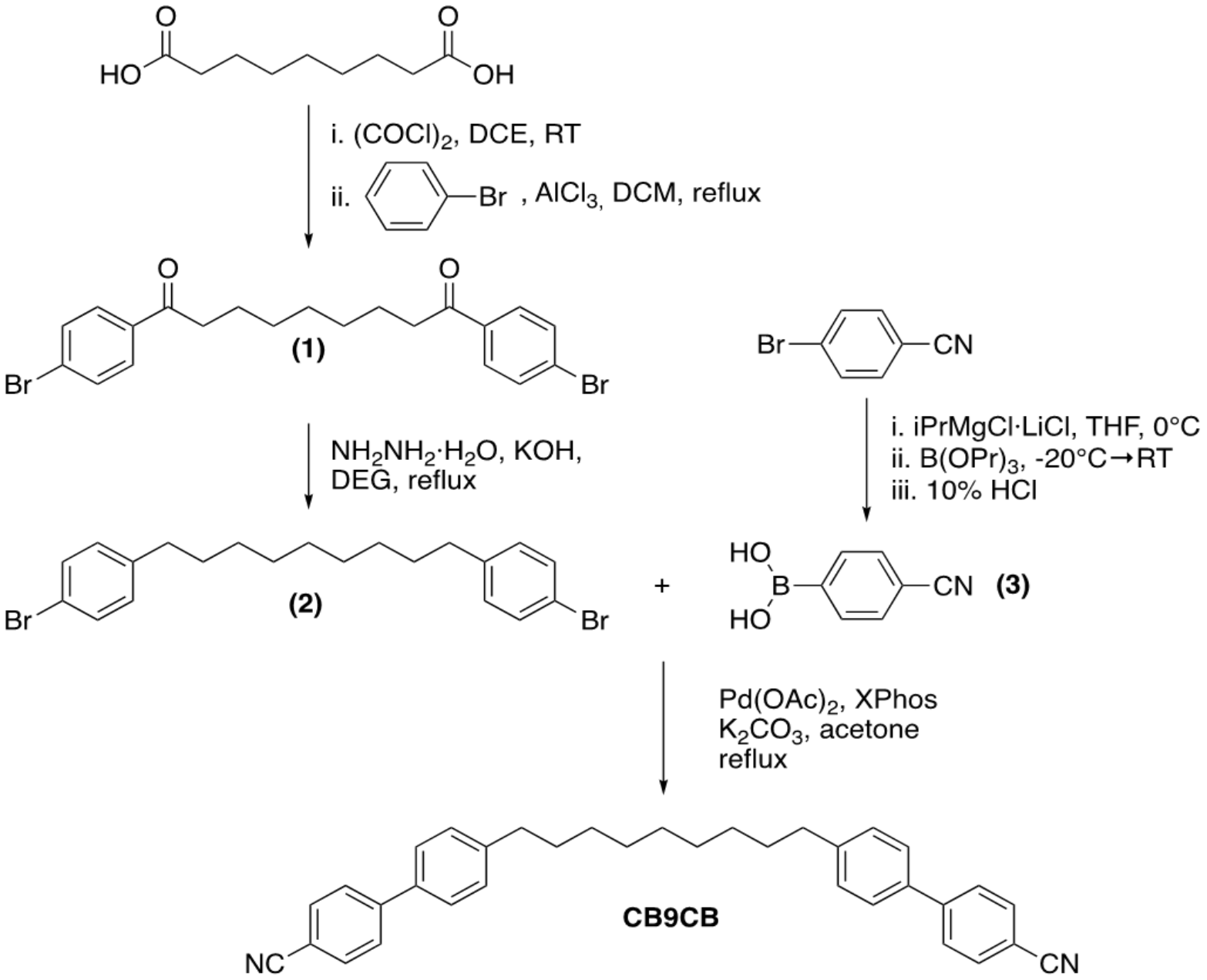}
\caption{Synthetic root of the compound CB9CB.
\label{fig:figure9}}
\end{figure}

The synthesis of CB9CB was performed as described in Fig.\ref{fig:figure9}.  Flexible spacer was introduced to the molecule in two steps. In the first step, Friedel-Crafts acylation of bromobenzene with azelaoyl chloride was performed and symmetrical diketone (1) was obtained. 
Semi-product (1) was then converted to corresponding $\alpha,\omega-$substituted nonane (2) via Wolff-Kishner reduction of diketone. Conversion of the 4-bromobenzonitrile to the (4-cyanophenyl)boronic acid (3) was optimised for magnesium-bromine exchange using lithium chloride complex of isopropylmagnesium chloride solution in THF, so called Turbo-Grignard reagent. The optimal temperature of the Mg-Br exchange was found to be within 0 and 10$^\circ$C. In the final step, CB9CB was obtained via Suzuki-Miyaura cross-coupling reaction between reagents (2) and (3) using palladium acetate and XPhos as the catalytic system. Final product was purified by double recrystallization and column chromatography. \\

\noindent{\bf Synthesis:}\\
\noindent\textit{1,9-bis(4-bromophenyl)nonane-1,9-dione (1)}:\\

 i. To a stirred solution of azelaic acid (37.6 g; 0.2 mol) in 1,2-dichloroethane (1 dm\textsuperscript{3}) oxalic chloride (36.7 cm\textsuperscript{3}; 0.42 mol) was added with a catalytic amount of N,N-dimethylformamide. The solution was stirred at room temperature for 1 day until full consumption of azelaic acid. The excess of oxalic chloride and 1,2-dichloroethane were removed under reduced pressure. Crude azelaoyl chloride was taken to the Friedel-Crafts acylation.
 
ii. To a stirred solution of AlCl\textsubscript{3} (52.13 g; 0.39 mol) in DCM (500 cm\textsuperscript{3}) azelaoyl chloride was added dropwise. During addition, the temperature went up by 5 $^\circ$C. The mixture was stirred for 2 hours until full consumption of aluminium trichloride. Bromobenzene (62.8 g; 0.4 mol) was then added drop-wise. The solution was refluxed for 16 hours. The reaction mixture was then poured out on ice water. Layers were separated, the organic layer was washed 4 times with water, dried over MgSO4 and concentrated under vacuum. The residue was recrystallized from acetone to yield 50 g (0.1073 mol, 54\%) of product 1. 
m.p. = 131$^\circ$C MS(EI) m/z: 446, 385, 267, 198, 183, 155\\

\noindent\textit{1,9-bis(4-bromophenyl)nonane (2)}:\\

The solution of 1,9-bis(4-bromophenyl)nonane-1,9-dione (42.8 g; 0.092 mol) and 80\% hydrazine hydrate (22.96 g; 0.367 mol) in diethylene glycol (250 cm\textsuperscript{3}) was stirred at 130 $^\circ$C for 1 hour. The mixture was cooled down to 120 $^{\circ}$C, KOH (29.8 g; 0.533 mol) was added, and the mixture was stirred at 130 $^{\circ}$C for 2 hours. Volatile compounds were then removed, and the mixture was kept at 200$^{\circ}$C for one hour. After cooling to the room temperature, the residue was washed three times with water to remove DEG and dissolved in dichloromethane. The organic layer was then washed three times with water, acidified with 10\% HCl and washed five times with water. The organic layer was dried and filtered through silica gel. DCM was evaporated and residue was recrystallized from anhydrous ethanol and acetone.

Obtained crystals were dissolved in hexane, filtered through silica gel and hexane was evaporated. The residue was recrystallized from ethanol and acetone to yield 26.6 g (0.0607 mol; 66\%) of product 2. m.p. = 30$^\circ$C MS(EI) m/z: 438, 247, 169, 91\\

\noindent\textit{ (4-cyanophenyl)boronic acid (3)}:\\

The reaction was carried out under N\textsubscript{2} atmosphere.
To a stirred mixture of 4-bromobenzonitrile (13.65 g; 0.075 mol) in THF iPrMgCl LiCl (73 cm\textsuperscript{3} of 1.04 M solution; 0.075 mol) was added dropwise at 0$^{\circ}$C. The reaction mixture was then stirred at 0$^{\circ}$C for 2 hours. After cooling to -20$^{\circ}$C, tripropyl borate (15.96 g; 0.085 mol) was added dropwise and the reaction mixture was allowed to reach room temperature. After hydrolysis (with 10\% HCl at pH=3) organic solvents were evaporated and crude product was recrystallized twice from water to yield 9.1 g (0.0619 mol; 83\%) of product 3. MS(EI) m/z: 187, 157, 129 (mass spectrum of the derived boronic ester).\\\\

\noindent\textit{$\alpha,\omega$-bis(4,4$'$-cyanobiphenyl) nonane CB9CB}:\\

The reaction was carried out under N\textsubscript{2} atmosphere.
The solution of 1,9-bis(4-bromophenyl)nonane (10 g; 0.02283 mol), (4-cyanophenyl)boronic acid (6.3 g; 0.04556 mol) and potassium carbonate (18.9 g; 0.137 mol) in 160ml of acetone/water mixture (1:1 v/v) was refluxed for 0.5 h. Then Pd(OAc)2 and XPhos (Pd 0.5 mol\% and 1 mol\% of the ligand) were added, and the reaction was refluxed for two hours. The reaction mixture was poured into water, and the residue was filtered off. The crude product was recrystallized from ethanol and acetone. Obtained crystals were dissolved in DCM, filtered through silica gel, and DCM was evaporated. The residue was recrystallized from ethanol and acetone to yield 10 g (0.0207 mol; 91\%) of CB9CB. MS(EI) m/z: 482, 207, 192, 165.


\begin{thebibliography}{99}


\bibitem{larson} R. G. Larson, \textit{The structure and rheology of complex fluids}, Oxford University Press, New York (1999).

\bibitem{dit} T. Divoux, M. A. Fardin, S. Manneville and S. Lerouge,  
\textcolor {blue} {Annu. Rev. Fluid Mech., {\bf 48}, 81 (2016)}.

\bibitem{iitm}J. M. Krishnan, A. P. Deshpande and P. B. Sunil Kumar, \textit{Rheology of complex fluids}, Springer, New York Dordrecht Heidelberg London (2010).

\bibitem{fielding} S. M. Fielding, \textcolor {blue} {Soft Matter \textbf{3}, 1262 (2007)}. 

\bibitem{pg} de Gennes, P. G. The Physics of Liquid Crystals; Oxford University Press: Oxford, England, (1974).

\bibitem{ch} S. Chandrasekhar, Liquid Crystals, Cambridge University Press, Cambridge, England (1992). 

\bibitem{blinov} L. M. Blinov and V. G. Chigrinov, \textit{Electrooptic effects in liquid crystal materials} Springer-Verlag, Berlin (1994).

\bibitem{safina}C. R. Safinya, E. B. Sirota, and R. J. Plano, \textcolor {blue}{Phys. Rev. Lett.,  \textbf{66}, 1986 (1991)}.

\bibitem{cates}O. Henrich, K. Stratford, D. Marenduzzo, P. V. Coveney and M. E. Cates, \textcolor {blue} {Soft Matter \textbf{12}, 3817 (2012)}.

\bibitem{sd}  J. Ananthaiah, R. Sahoo, M. V. Rasna, and S. Dhara, \textcolor {blue} {Phys. Rev. E \textbf{89}, 022510 (2014)}.

\bibitem{nagita} K. Negita, M. Inoue, and S. Kondo, \textcolor {blue} {Phys. Rev. E \textbf{74}, 051708 (2006)}.

\bibitem{mr}M. R., Alcantara. J. Vanin, \textcolor {blue} {Colloids and Surfaces A: Physicochemical and Engineering Aspects \textbf{97}(2) 151 (1995)}.

\bibitem{rm} R. Mezzenga, C. Meyer, C. Servais, A. I. Romoscanu, L. Sagalowicz, and R. C. Hayward, \textcolor {blue} {Langmuir, \textbf{21}, 3322 (2005)}.

\bibitem{ad} W. H. Han and A. D. Rey, \textcolor {blue} {J. Rheol. \textbf{39}, 301 (1995)}.


\bibitem{id} I. Dozov, \textcolor {blue} {Europhys. Lett. \textbf{56}, 247 (2001)}.

\bibitem{vp}V. P. Panov, M. Nagaraj, J. K. Vij, Yu. P. Panarin, A. Kohlmeier, M. G. Tamba, R. A. Lewis, and G. H. Mehl \textcolor{blue}{Phys. Rev. Lett., \textbf {105}, 167801 (2010).} 


\bibitem{vb}V. Borshch, Y.-K. Kim, J. Xiang, M Gao, A Jakli, V. P. Panov, J. K. Vij, C. T. Imrie, M. G. Tamba, G. H. Mehl, and O. D. Lavrentovich,  \textcolor {blue} {Nat. Commun., \textbf {4}, 2635 (2013).}

\bibitem{mco} M. Copic,  \textcolor {blue} { Proc. Natl. Acad. Sci. USA,  \textbf {110},  15931 (2013).}


\bibitem{chd} D. Chen, J. H. Porada, J. B. Hooper, A. Klittnick, Y. Shena, M. R. Tuchbanda, E. Korblova, D. Bedrov, D. M. Walba, M. A. Glaser, J. E. Maclennana, and N. A. Clark, \textcolor {blue} { Proc. Natl. Acad. Sci. USA,  \textbf {110}(40),  15855 (2013).}

\bibitem{mcl} C. Meyer, G. R. Luckhurst, and I.  Dozov,  \textcolor {blue} {Phys. Rev. Lett., \textbf {111}, 067801 (2013).}

\bibitem{prx} Z. Parsouzi, S. M. Shamid, V. Borshch, K. Challa, A. R. Baldwin, M. G. Tamba, C. Welch, G. H. Mehl, J. T. Gleeson, A. Jakli, O. D. Lavrentovich, D. W. Allender, J. V. Selinger and S. Sprunt, \textcolor {blue} {Phys. Rev. X \textbf {6}, 021041 (2016)}.

\bibitem{oleg1} Z. Parsouzi, S. A. Pardaev, C. Welch, Z. Ahmed, G. H. Mehl, A. R. Baldwin, J. T. Gleeson, O. D. Lavrentovich, D. W. Allender, J. V. Selinger, A. Jakli and S. Sprunt, \textcolor {blue} {Phys. Chem. Chem. Phys.,  \textbf{18}, 31645 (2016)}.

\bibitem{noel} C. Zhu, M. R. Tuchband, A. Young, Min Shuai, A. Scarbrough, D. M. Walba, J. E. Maclennan, C. Wang, A. Hexemer, and N. A. Clark, \textcolor {blue} {Phys. Rev. Lett. \textbf {116}, 147803 (2016)}.

\bibitem{cm} C. Meyer and I. Dozov, \textcolor {blue} {Soft Matter \textbf{12}, 574 (2016)}.



\bibitem{sms} S. M. Salili, C. Kim, S. Sprunt, J. T. Gleeson, O. Parri and A. Jakli,  \textcolor {blue} {RSC Adv.,  \textbf{4}, 57419 (2014)}.

\bibitem{prm} M. Praveen Kumar, P. Kula, and S. Dhara, \textcolor {blue} {Phys. Rev. Materials, \textbf {4}, 115601 (2020)}.

\bibitem{da} D.A. Paterson, J. P. Abberley, W. TA. Harrison, J. MD Storey and C. T. Imrie, \textcolor {blue} {Liq. Cryst. \textbf{44}, 127 (2017)}.

\bibitem{rg}R. Ganapathy and A. K.Sood, \textcolor {blue} {Langmuir, \textbf{22}, 11016 (2006).}

\bibitem{OD} O. Diat, D. Roux, F. Nallet, \textcolor {blue} {J. Phys. II France, \textbf {3}, 1427 (1993).} 

\bibitem{dr} D. Roux, F. Nallet, and O. Diat \textcolor {blue} {Europhys Lett, \textbf {24}, 53 (1993).}

\bibitem{pza} P. Panizza, P. Archambault and D. Roux, \textcolor{blue} {J. Phys. II France, \textbf {5}, 303 (1995)}


\bibitem{rb} R. Bandyopadhyay, A. K. Sood, \textcolor {blue} {Langmuir, \textbf {19}, 3121 (2003).}

\bibitem{jb} J. Berret, D. C. Roux and G. Porte, \textcolor {blue} {J. Phys. II France, \textbf {4}, 1261 (1994).} 


 

\bibitem{jfb}J. F. Berret, \textcolor {blue} {Langmuir, \textbf{13}, 2227 (1997).}

\bibitem{pramana} A. K. Sood and R. Ganapathy, \textcolor {blue} {Pramana, \textbf{67}, 33 (2006).}


\bibitem{smf} S. M. Fielding and P. D. Olmsted, \textcolor {blue} {Eur. Phys. J. E , \textbf{11}, 65 (2003).} 


\bibitem{gp}  G. Porte, J-F. Berret and J. L. Harden, \textcolor {blue} {J. Phys. II France, \textbf{7}, 459 (1997).} 

\bibitem{rbprl} R. Bandyopadhyay, G. Basappa and A. K. Sood, \textcolor {blue} {Phys. Rev. Lett., \textbf {84}, 2022 (2000).}

\bibitem{smfprl} S. M. Fielding and P. D. Olmsted, \textcolor {blue} {Phys. Rev. Lett., \textbf{92}, 084502 (2004).} 

\bibitem{zhu} C. Zhu, M. R. Tuchband, A. Young, M. Shuai, A. Scarbrough, D. M. Walba, J. E. Maclennan, C. Wang, A. Hexemer, and N. A. Clark, \textcolor {blue} {Phys. Rev. Lett. \textbf{116}, 147803 (2016)}.



\bibitem{rgprl} R. Ganapathy and A. K. Sood, \textcolor {blue} {Phys. Rev. Lett., \textbf {96}, 108301 (2006).}



\bibitem{srprl} B. Chakrabarti, M. Das, C. Dasgupta, S. Ramaswamy and A. K. Sood, \textcolor {blue} {Phys. Rev. Lett., \textbf {92}, 055501 (2004).}

\bibitem{cd1} R. Mandal, B. Chakrabarti, D. Chakrabarti,  C. Dasgupta,  \textcolor {blue} {J. Phys.: Condens. Matter, \textbf {32}, 134002 (2020).}


\bibitem{mdas} M. Das, B. Chakrabarti, C. Dasgupta, S. Ramaswamy, and A. K. Sood \textcolor {blue} {Phys. Rev. E \textbf{71}, 021707 (2005)}. 

\bibitem{dc1} D. Chakrbarti, C. Dasgupta, and A. K. Sood \textcolor {blue} {Phys. Rev. E \textbf{82}, 065301R (2010)}. 


\bibitem{rgm} G. Rien\"{a}cker, M. Kr\"{o}ger and S. Hess, \textcolor {blue} {Phys. Rev. E \textbf{66}, 040702R (2002)}. 





























\end {thebibliography}
\end{document}